\documentclass[a4paper,11pt]{article}
\usepackage{jheppub} 
\usepackage{lineno}
\usepackage{amsthm}

\usepackage{tikz-cd}
\usetikzlibrary{decorations.pathmorphing}
\usepackage{setspace}
\usepackage{float}

\usepackage{cancel}

\newcommand{\vol}{\mathrm{vol}}

\usepackage[dvipsnames]{xcolor}
\usepackage{soul}

\title{Einstein--\ae{ther} Elliptic Charges and the First Law of Asymptotically AdS Black Holes}

\author[a, b, c]{W. Arata,}
\author[d]{M. Ismail,}
\author[a, b, c]{S. Liberati,}
\author[d,1]{L. Martin\note{Corresponding author},}
\author[d]{D. Mattingly,}
\author[a,b,c]{and G. Neri,}

\affiliation[a]{SISSA, via Bonomea 265, 34136 Trieste, Italy}
\affiliation[b]{INFN (Sez.~Trieste), via Valerio 2, 34127 Trieste, Italy}
\affiliation[c]{IFPU, via Beirut 2, 34014 Trieste, Italy}
\affiliation[d]{University of New Hampshire, Durham, NH 03824, USA}

\emailAdd{warata@sissa.it}
\emailAdd{MoustafaKamel.Ismail@unh.edu}
\emailAdd{liberati@sissa.it}
\emailAdd{Luke.Martin@unh.edu}
\emailAdd{David.Mattingly@unh.edu}
\emailAdd{gneri@sissa.it}

\abstract{We investigate the thermodynamic role of asymptotic \ae{ther} alignment for universal horizons in Einstein--\ae{ther} theory. In the static, spherically symmetric, asymptotically AdS sector with $c_{14}=0$, the known first law for universal horizons contains an additional term whenever the \ae{ther} is misaligned with the timelike Killing vector at infinity. While this term has recently been interpreted in Ho\v{r}ava--Lifshitz gravity as the contribution of an elliptic charge associated with khronon reparameterizations, no corresponding explanation was available in Einstein--\ae{ther} theory. We show that, in the same sector, Einstein--\ae{ther} theory possesses a previously unidentified symmetry of the reduced action, generated by infinitesimal transformations of the form $\delta u^a=f a^a$, where $a^a$ is the \ae{ther} acceleration and $f$ obeys an elliptic constraint. We derive the associated current and charge, and show that the aligned limit is naturally interpreted as the ensemble in which this \ae{ther}-charge contribution vanishes. This provides the Einstein--\ae{ther} counterpart of the elliptic-charge mechanism in Ho\v{r}ava--Lifshitz gravity and clarifies the thermodynamic significance of asymptotic \ae{ther} alignment.}

\begin{document}
\maketitle
\flushbottom

\section{Introduction}
\label{sec:intro}

Black holes have long provided one of the most fertile grounds for probing the structure of candidate theories of quantum gravity. Black-hole thermodynamics is understood to arise as a thermodynamic limit of an underlying quantum theory, and its failure often signals that such a fundamental quantum description is incomplete or ill-behaved. An important feature of black-hole thermodynamics is its sensitivity to the symmetry structure of the underlying gravitational theory. Through Noether's theorem,
horizon entropy and the first law governing its variations are tied to the symmetries of the theory.
The relevant notion of horizon is determined by the causal structure
available to the propagating degrees of freedom. Thus, when the symmetry or
causal structure of a gravitational theory is modified, the thermodynamics of
its black holes may change as well. Understanding these changes can illuminate
both the internal consistency and the physical content of the theory.

Einstein--\ae ther theory~\cite{Jacobson:2000xp} is a generally covariant
modification of general relativity in which the metric is coupled to a dynamical unit
timelike vector field, the \ae ther. The \ae ther selects a preferred local
frame, breaking local Lorentz invariance while preserving spacetime
diffeomorphism invariance. In addition to the usual spin-2 graviton, the theory contains spin-1 and spin-0 modes associated with the \ae ther.  These modes generally
propagate at speeds different from one another and from the metric speed of
light. Consequently, the metric light cone alone does
not determine the causal structure relevant for all degrees of freedom. This has immediate consequences for black holes. If the theory
is extended na{\"i}vely into the ultraviolet while the different mode speeds
remain finite, a black-hole spacetime may contain multiple horizons, one for
each propagating sector. This leads to thermodynamic pathologies, including the possibility of perpetual motion machines~\cite{Dubovsky:2006vk}. This is generally viewed as a sign that Einstein--\ae ther theory is an infrared effective theory whose ultraviolet completion needs to modify the high-energy causal structure.

Ho\v{r}ava--Lifshitz gravity~\cite{Horava_Lifshitz,Horava_Membranes} provides a closely related framework in which this ultraviolet behavior is built in from the outset~\cite{Blas:2025ejp}. The theory introduces a preferred foliation, defined by a scalar
field called the khronon, and its high-energy Lifshitz scaling allows
excitations to propagate with arbitrarily large speeds relative to that
foliation. Nevertheless, the khronon still defines a causal ordering. Black-hole solutions can therefore contain universal horizons~\cite{Barausse:2011pu}, which trap even infinitely fast signals because every
future-directed trajectory must move forward in khronon time. Universal
horizons are thus the natural causal boundaries for black holes in theories
with a preferred foliation. In Ho\v{r}ava--Lifshitz gravity, they have been shown to possess a well-defined thermodynamics~\cite{Mechanics,Berglund:2012fk,DelPorro:2022vqi,DelPorro:2023lbv} and to evade the problems encountered in Einstein--\ae ther theory~\cite{Benkel:2018abt}.

The connection to Einstein--\ae ther theory becomes clear in the twist-free, static, spherically symmetric sector. In this sector, despite their different ultraviolet interpretations and horizon structures, the solution space of Einstein--\ae ther theory has been proven to be equivalent to that of the infrared limit
of Ho\v{r}ava--Lifshitz gravity~\cite{Max_Sym_UH}. This is practically useful since results established in one theory can be translated into the other, and the two theories can be studied in tandem. In particular, both theories admit the same family of static, spherically symmetric solutions \textit{with} universal horizons.
The mechanics of these solutions---the Smarr formula, the mass formula, and the first law---can therefore be investigated equally well in either framework.

A difficulty has persisted, however, in the thermodynamics of asymptotically AdS universal horizons. While the first law is well-behaved in the asymptotically flat case when the {\ae}ther is aligned with the timelike Killing vector at infinity \cite{Mechanics}, attempts to formulate a first law for asymptotically AdS universal horizons have consistently produced an additional term with no clear thermodynamic interpretation. Specifically, the variation of the properly background-subtracted Komar mass yields~\cite{Max_Sym_UH,Martin2024-tb}
\begin{equation}\label{eq:first law}
    \delta M = \frac{\kappa}{8\pi G_\textnormal{\ae}} \delta A
             + \frac{\mathcal{V}}{8\pi G_\textnormal{\ae}} \delta \Lambda
             + (\text{extra}),
\end{equation}
where $G_\textnormal{\ae} = G_N \, (1 - c_{14}/2)^{-1}$ is the effective Newton constant. The first term on the right-hand side is the expected entropy contribution,\footnote{See Sec.~\ref{subsec:UH thermo} for its quite non-trivial interpretation} while the second is the enthalpy term familiar from AdS black-hole thermodynamics \cite{Traschen}. The extra term vanishes only when the \ae ther aligns asymptotically with the timelike Killing vector. 
Thus, only the asymptotic alignment appears to select an equilibrium ensemble appropriate for a coherent first law. While heuristic arguments for this occurrence were discussed in the literature, the physical \emph{mechanism} behind this selection has remained obscure.

Recently, in Ho\v{r}ava--Lifshitz gravity, this mechanism was finally identified in Ref.~\cite{Martin2024-tb}. In this case, the theory is invariant under khronon reparameterizations $T\to\tilde T(T)$, which relabels the preferred leaves without changing the foliation itself.
This symmetry is neither an ordinary global symmetry nor a fully local gauge symmetry. Rather, it is local in the preferred time coordinate but global on each leaf. Its associated Noether current describes the flow of an \emph{elliptic charge}, denoted $Q_T$. Conservation of this charge is equivalent to the vanishing of a spatial flux through the boundary of each preferred-time slice. 

For the asymptotically AdS universal-horizon solutions, this flux is controlled by a parameter, $\ell_s$, which characterizes the asymptotic misalignment between the \ae{ther} and the timelike Killing vector. The flux vanishes in the limit of exact alignment, 
$\ell_s\to\infty$, and it is non-zero for any finite $\ell_s$. In the aligned case, the boundary contribution associated with the elliptic charge vanishes, and the first law reduces to the entropy--enthalpy form. The extra term in Eq.~\eqref{eq:first law} can therefore be interpreted, in the Ho\v{r}ava--Lifshitz formulation, as the boundary contribution associated with
the elliptic charge.

No analogous explanation has been available thus far in Einstein--\ae ther theory. Although the same universal-horizon solutions and the same first-law obstruction appear in the two formulations, the reparameterization symmetry responsible for the elliptic charge
in Ho\v{r}ava--Lifshitz gravity is trivial in Einstein--\ae ther theory. To see this, recall that the \ae ther one-form is the fundamental field in Einstein--\ae ther theory, whereas the khronon is fundamental in Ho\v{r}ava--Lifshitz gravity. In the twist-free sector, the two are connected by writing the \ae ther as
\begin{equation}
    u_a=-N \nabla_a T,
\end{equation}
where $N$ ensures the unit normalization. Under a khronon reparameterization $T\to\tilde T(T)$, the lapse transforms so that the one-form $u_a$ is
unchanged. Thus, the Ho\v{r}ava--Lifshitz elliptic charge has no immediate counterpart when the \ae ther is taken as the fundamental field. This
leaves a conceptual gap: the same asymptotically AdS universal-horizon first law
and the same alignment condition arise in Einstein--\ae ther theory, but the
charge-based mechanism selecting the aligned ensemble has not yet been
identified in Einstein--\ae ther theory.

In this paper, we close this gap by identifying the Einstein--\ae ther mechanism. We show that, in the
$c_{14}=0$ sector and under the assumption of spherical symmetry,
Einstein--\ae ther theory possesses a previously unidentified symmetry of the
action. The infinitesimal symmetry transformation takes the form
\begin{equation}
    u_a \mapsto u_a + f a_a,
    \label{eq:aether_symmetry}
\end{equation}
where $a_a = u^b \nabla_b u_a$ is the \ae ther acceleration. Since $a_a$ is
orthogonal to $u_a$, this transformation preserves the unit-norm constraint at
linear order. The function $f$ is restricted by the elliptic equation
\begin{equation}
    \nabla_a(f a^a)=0 .
\end{equation}
We derive the associated Noether current and define the corresponding
\emph{\ae ther charge} $Q_f$. Although this charge vanishes on hypersurfaces
orthogonal to the \ae ther, its spatial flux is non-trivial and supplies the
boundary contribution relevant to the asymptotically AdS first law.

We then show that the boundary variation associated with $Q_f$ vanishes precisely in the aligned limit $\ell_s\to\infty$. Asymptotic alignment
therefore acquires a charge-based interpretation directly within
Einstein--\ae ther theory. It is the condition under which the \ae ther-charge
contribution at the asymptotic boundary is absent.  For finite misalignment,
the same boundary contribution accounts for the extra term in
Eq.~\eqref{eq:first law}.  The \ae ther charge thus provides the
Einstein--\ae ther counterpart of the elliptic-charge mechanism previously
identified in Ho\v{r}ava--Lifshitz gravity.

The structure of the paper is as follows.  In Sec.~\ref{sec:background} we
review the Einstein--\ae ther action, its relation to the infrared limit of
Ho\v{r}ava--Lifshitz gravity, the static spherically symmetric asymptotically AdS universal-horizon solutions in the $c_{14}=0$ sector, and the current status of the first law for these solutions. In Sec.~\ref{sec:symmetry} we identify the new \ae ther symmetry, derive the elliptic condition on $f$, construct the corresponding current, define the charge $Q_f$, and show through the Hamiltonian analysis that
$Q_f$ generates the symmetry transformation. In Sec.~\ref{sec:ensemble} we show that the flux associated with changes in $Q_f$ vanishes if and only if the \ae ther is aligned at infinity, thereby providing the physical underpinning of the first law as a choice of ensemble. We conclude in
Sec.~\ref{sec:summary} by discussing the implications of these results, their relationship to the Ho\v{r}ava--Lifshitz analysis of Ref.~\cite{Martin2024-tb}, and directions for future work.

\section{Background}\label{sec:background}

\subsection{Einstein--\ae ther theory}\label{subsec:Aether Intro}

Einstein--\ae ther theory is a generally covariant theory of gravity in which the metric $g_{ab}$ is supplemented by a dynamically preferred timelike $1$-form,\footnote{One could equivalently take the \ae ther to be fundamentally a vector field. For the purposes of this paper, however, it is convenient to regard it as a $1$-form.} called the \ae ther~\cite{aether_waves}. We denote the components of this $1$-form by $u_a$, and those of the dual vector field by $u^a$. The theory is described by the action
\begin{equation}\label{eq:aether action}
S=\frac{1}{16\pi G_{\textnormal{\ae}}}\int_{\mathcal{M}}\vol_\mathcal{M}\left[-\frac{6c_{\mathrm{cc}}}{\ell^2}+R+\mathcal{L}_{\textnormal{\ae}}+\mathcal{L}_{\textnormal{\ae}}^{\mathrm{(con)}}\right].
\end{equation}
Here $R$ is the Ricci scalar of $g_{ab}$, while the first term is the bare cosmological-constant contribution. The parameter $c_{\mathrm{cc}}=0,\pm1$ fixes the sign of the bare cosmological constant, and $\ell$ fixes the scale.
The term $\mathcal{L}_{\textnormal{\ae}}^{(\mathrm{con})}$ imposes the unit-norm constraint on the \ae ther,
\begin{equation}\label{eq:L con}
\mathcal{L}_{\textnormal{\ae}}^{\mathrm{(con)}}=\lambda_{\textnormal{\ae}}(u^a u_a +1),
\end{equation}
where $\lambda_{\textnormal{\ae}}$ is a Lagrange multiplier field. The remaining term, $\mathcal{L}_{\textnormal{\ae}}$, contains the derivative self-interactions of the \ae ther. Truncating to terms quadratic in $\nabla_a u^b$, one writes 
\begin{equation}\label{eq:L ae}
\mathcal{L}_{\textnormal{\ae}} = - Z^{ab}_{\;\;\;cd}(\nabla_a u^c)(\nabla_b u^d) \,,
\end{equation}
with
\begin{equation}\label{eq:Z tensor}
Z^{ab}_{\;\;\;cd}=c_1g^{ab}g_{cd}+c_2\delta^a_{\;c}\delta^b_{\;d}+c_3\delta^a_{\;d}\delta^b_{\;c}-c_4u^au^bg_{cd} \,.
\end{equation}
The four constants $c_i$ are dimensionless running couplings. It will be useful to introduce the combinations
\begin{equation}\label{eq:couplings}
\begin{gathered}
c_{13}=c_1+c_3 \,,\hspace{7mm}\overline{c}_{13}=c_1-c_3 \,,\hspace{7mm}c_{14}=c_1+c_4 \,, \hspace{5mm}\vspace{5mm}\\
c_{123}=c_1+c_2+c_3 \,, \hspace{7mm}c_\ell=\frac{2+c_{13}+3c_2}{2} \,.
\end{gathered}
\end{equation}
For twist-free \ae ther configurations, namely those satisfying $u\wedge \mathrm{d} u=0$, solutions of Einstein--\ae ther theory depend only on the combinations $c_{13}$, $c_2$, and $c_{14}$ in the action~\cite{extended_horava}. To avoid pathologies such as negative-energy modes and naked singularities, one additionally imposes the bounds~\cite{Jacobson:2007fh}
\begin{equation}\label{eq:coupling bounds}
c_{13}<1, \qquad 0\leq c_{14}<2, \qquad c_{123}\geq 0.
\end{equation}

It is also useful to express the \ae ther derivative $\nabla_a u_b$ in terms of the standard kinematical quantities of a timelike congruence. Defining the spatial projector 
\begin{align}
    h_{ab}=g_{ab}+u_a u_b,
\end{align}
one has
\begin{equation}\label{eq:du decomp}
\nabla_a u_b
=
- u_a a_b
+ \frac{1}{3}\vartheta h_{ab}
+ \sigma_{ab}
+ \omega_{ab},
\end{equation}
where
\begin{alignat}{2}
a^a &= u^b \nabla_b u^a , \quad &&(\text{acceleration})\label{eq:accel defn}\\
\vartheta &= h^{ab} \nabla_a u_b , \quad &&(\text{expansion})\label{eq:expansion defn}\\
\sigma_{ab} &= h_a{}^c h_b{}^d \nabla_{(c}u_{d)}
- \frac{1}{3}\vartheta h_{ab} , \quad &&(\text{shear})\label{eq:shear defn}\\
\omega_{ab} &= h_a{}^c h_b{}^d \nabla_{[c}u_{d]} . \quad &&(\text{twist})\label{eq:twist defn}
\end{alignat}
In terms of these quantities, the \ae ther Lagrangian becomes
\begin{equation}\label{eq:Lae twist etc}
\mathcal{L}_{\textnormal{\ae}}=- c_\vartheta \vartheta^2
- c_\sigma \sigma_{ab}\sigma^{ab}
- c_\omega \omega_{ab}\omega^{ab}
+ c_a a^b a_b.
\end{equation}
The couplings appearing in this form are related to those in Eqs.~\eqref{eq:Z tensor} and~\eqref{eq:couplings} by
\begin{equation}\label{eq:couplings dictionary}
c_\vartheta=\dfrac{c_{13}+3c_2}{3}, \qquad c_\sigma=c_{13}, \qquad c_\omega=\overline{c}_{13}, \qquad c_a=c_{14}.
\end{equation}

The field equations follow by demanding that the action in Eq.~\eqref{eq:aether action} be stationary, up to possible boundary terms, under variations of the metric, the \ae ther, and the Lagrange multiplier field. Following the notation of Refs.~\cite{Mechanics,Max_Sym_UH}, the field equations are
\begin{equation}\label{eq:eoms}
\mathcal{G}_{ab}=-\frac{3c_{\mathrm{cc}}}{\ell^2}g_{ab}+\mathcal{T}_{ab}, \qquad \text{\AE}_a+\lambda_\textnormal{\ae}u_a=0, \qquad u^a u_a +1 = 0,
\end{equation}
where $\mathcal{G}_{ab}$ is the Einstein tensor, and
\begin{align}
\mathcal{T}_{ab} &= \lambda_{\text{\ae}}u_au_b+c_4a_aa_b-\frac{1}{2}g_{ab}Y^c_{\;\;d}\nabla_cu^d+\nabla_c X^c_{\;\;ab} \notag\\
&~~~+c_1\left[(\nabla_a u_c)(\nabla_b u^c)-(\nabla^c u_a)(\nabla_c u_b)\right],\label{eq:eom specifics-Stress-Tensor}\\
\text{\AE}_a & = c_4(\nabla_a u^b)a_b + \nabla_b Y^b_{\;\;a},\label{eq:eom specifics-Aether}\\
Y^a_{\;\;b}&=Z^{ac}_{\;\;bd}\nabla_c u^d,\label{eq:eom specifics-Y}\\
X^c_{\;\;ab}&=Y^c_{\;\;(a}u_{b)}-u_{(a}Y_{b)}^{\;\;c}+u^c Y_{(ab)}.\label{eq:eom specifics-X}
\end{align}
The tensors $X^c_{\;\;ab}$ and $Y^a_{\;\;b}$ are auxiliary quantities that compactly organize the metric and \ae ther variations.\footnote{We use round brackets for symmetrization and square brackets for antisymmetrization. These operations include the conventional factor of $1/n!$, where $n$ is the number of indices being symmetrized or antisymmetrized.}

A noteworthy feature of Einstein--\ae ther theory is the existence of propagating spin-0, spin-1, and spin-2 modes, the speeds of which in the \ae ther frame are respectively~\cite{aether_waves}
\begin{align}
s_0^2&=\frac{c_{123}}{c_{14}(1-c_{13})c_\ell}\left(1-\frac{c_{14}}{2}\right),\label{eq:mode-0 speed}\\
s_1^2&=\frac{c_{13}+\overline{c}_{13}-c_{13}\overline{c}_{13}}{2(1-c_{13})c_{14}},\label{eq:mode-1 speed}\\
s_2^2&=\frac{1}{1-c_{13}}.\label{eq:mode-2 speed}
\end{align}
Consequently, the relevant causal structure is not determined by the metric light cone, but rather by the fastest (possibly superluminal) propagating mode, whose causal cone is described by the corresponding ``speed-$s$ metric''~\cite{UHOG,Causality}
\begin{equation}\label{eq:speed s metric}
g^{(s)}_{ab}=g_{ab}-(s_i^2-1)u_au_b
\end{equation}
where $i=0,1,2$ labels the spin of the mode under consideration. Equations~\eqref{eq:mode-0 speed}--\eqref{eq:mode-2 speed} and~\eqref{eq:coupling bounds} show, in particular, that the spin-0 and spin-1 speeds are not generally bounded from above.  In the special case $c_{14}=c_a=0$, both the spin-0 and spin-1 mode speeds diverge in the \ae ther frame, but the latter is absent in the spherically symmetric configurations considered here~\cite{aether_waves}. 
\subsection{Connection between Einstein--\ae ther theory and Ho\v{r}ava--Lifshitz gravity}

Before turning to causal horizons in Einstein--\ae ther theory, it is useful to recall the connection between Einstein--\ae ther theory and Ho\v{r}ava--Lifshitz gravity. Up to boundary terms, the Einstein--\ae ther action is equivalent to the twist-free infrared limit of Ho\v{r}ava--Lifshitz gravity \cite{Status,extended_horava}. Ho\v{r}ava--Lifshitz gravity achieves power-counting renormalizability by introducing a preferred foliation of spacetime, thereby sacrificing Lorentz invariance at high energies \cite{Horava_Lifshitz,Horava_Membranes}. This relation to a candidate power-counting-renormalizable theory of quantum gravity makes Einstein--\ae ther theory worth studying in its own right. 

In the twist-free sector, the \ae ther is locally hypersurface orthogonal and can be written as $u=-N\,\mathrm{d}T$,
where $T$ is the khronon field, whose level sets define the preferred spacelike hypersurfaces $\Sigma$, and $N$ is the lapse. In the preferred frame, $T$ plays the role of time, and the theory is invariant under smooth (monotonic) time reparametrizations. The unit-norm condition fixes the lapse to be
\begin{equation}\label{eq:Lapse in T}
N=\frac{1}{\sqrt{-g^{ab}(\nabla_a T)(\nabla_b T)}}
\end{equation}

The infrared equivalence between the two theories should nevertheless be understood with care---the fundamental degrees of freedom are different. In the $T$-theory formulation of Ho\v{r}ava--Lifshitz gravity, the khronon $T$ is fundamental, whereas in Einstein--\ae ther theory the fundamental object is the unit timelike \ae ther field normal to the constant-$T$ slices. Because of the close relationship between the two theories, however, we will freely borrow notation and intuition from the $T$-theory formulation of Ho\v{r}ava--Lifshitz gravity into the Einstein--\ae ther discusstion whenever useful.

\subsection{Maximally symmetric universal horizons}\label{subsec:UH}

Because the propagation speeds of the modes in Einstein--\ae ther theory need not have an upper bound \cite{Causality}, the notion of a causal horizon is more subtle than in general relativity. In general relativity, one often studies globally hyperbolic spacetimes in which causal horizons coincide with bifurcate Killing horizons of the spacetime metric. In Einstein--\ae ther theory, however, a Killing horizon of $g_{ab}$ does not necessarily trap arbitrarily fast signals, and therefore cannot by itself define a causal horizon for the full theory. Instead, one may have a \emph{universal horizon}, which traps all future-directed causal curves as defined by the preferred \ae ther frame \cite{UHOG}.

A future-directed causal curve $\gamma$ in Einstein--\ae ther theory is defined as a curve whose tangent vector $t^a$ satisfies $(u\cdot t) \leq 0$~\cite{Causality}. A universal horizon must then be a hypersurface beyond which every future-directed causal curve, including curves associated with arbitrarily fast modes, is forced toward smaller radius. Since the preferred time is defined by the \ae ther, such a horizon is naturally characterized by the behavior of the \ae ther relative to the stationary Killing field $\chi$. Indeed, we will show that the \ae ther becomes orthogonal to the Killing vector field at a universal horizon, $(u \cdot \chi) =0$.

Universal horizons are most easily studied in static, spherically symmetric spacetimes. In this sector, the twist $u\wedge \mathrm{d}u$ vanishes automatically. We introduce the unit spacelike vector $s^a$ in the radial direction, chosen to be orthogonal to the \ae ther and aligned with the acceleration (which is always non-vanishing in universal-horizon solutions): 
\begin{equation}\label{eq:s defn}
a^a=(a\cdot s)s^a, \qquad (u\cdot s)=0, \qquad s^2=1.
\end{equation}
We work in ingoing Eddington--Finkelstein coordinates, which are regular away from the curvature singularity at $r=0$. The line element is written as
\begin{equation}\label{eq:EF line element}
ds^2=-e(r)\,\mathrm{d}v^2 +2F(r)\,\mathrm{d}r\,\mathrm{d}v+r^2\mathrm{d}\Omega_{2}^2,
\end{equation}
where $\mathrm{d}\Omega_{2}^2$ denotes the standard line element on the unit $2$-sphere. The timelike Killing vector is $\chi=\partial_v$ and thus $e(r)=-\chi \cdot \chi$. Using the orthonormal basis $\{u^a,s^a\}$ in the temporal-radial plane, the Killing vector decomposes as
\begin{equation}\label{eq:chi}
\chi^a=-(u\cdot \chi)u^a+(s\cdot \chi)s^a.
\end{equation}
It follows then that 
$e(r)=(u\cdot\chi)^2-(s\cdot\chi)^2.$
The \ae ther 1-form can then be written as~\cite{Max_Sym_UH}
\begin{equation}\label{eq:aether EF}
u = u_a \mathrm{d} x^a=(u\cdot\chi)\,\mathrm{d}v+\frac{F(r)\,\mathrm{d}r}{(s\cdot\chi)-(u\cdot\chi)}.
\end{equation}

To identify the relevant causal hypersurface, it is useful to introduce coordinates adapted to the preferred foliation defined by the \ae ther. In a gauge satisfying $\chi^a \nabla_a T=1$,
the lapse function, defined by $-N\mathrm{d}T=u_a \mathrm{d}x^a$, takes the form~\cite{Max_Sym_UH}
\begin{equation}\label{eq:lapse udotchi}
N=-(u\cdot \chi). 
\end{equation}
In this gauge, the \ae ther may be written as $u=(u\cdot\chi)\,\mathrm{d}T$, where
\begin{equation}\label{eq:dT EF}
\mathrm{d}T=\mathrm{d}v+\frac{F(r)\,\mathrm{d}r}{(u\cdot\chi)\left[(s\cdot\chi)-(u\cdot \chi)\right]}.
\end{equation}
Equation~\eqref{eq:dT EF} shows that the preferred-time slicing becomes singular when $(u\cdot\chi)=0$. This condition defines the universal horizon. Beyond this surface, every signal that moves forward in khronon time, i.e., $\mathrm{d}T>0$, is forced toward $r=0$, i.e., $\mathrm{d}r<0$.

Globally maximally symmetric solutions and spherically symmetric black-hole solutions with universal horizons were constructed in Ref.~\cite{Max_Sym_UH}. In this paper, we restrict attention to the asymptotically AdS solutions since, for the case of interest here where $c_{14}=c_a=0$, this is the relevant family of solutions. These solutions are usefully characterized by four mutually dependent parameters: $\ell_u$, $\ell_s$, $\ell$, and $\Lambda$. The parameters $\ell_u$ and $\ell_s$ respectively determine the asymptotic behavior of $(u\cdot\chi)$ and $(s\cdot\chi)$ as $r\to\infty$. The parameter $\ell$ is associated with the bare cosmological constant term in the action. Finally, the effective cosmological constant $\Lambda$ receives contributions from both the bare cosmological constant term and the \ae ther behavior at infinity.  These parameters are related via
\begin{equation}\label{eq:l relations}
\begin{aligned}
\frac{c_{\mathrm{cc}}}{\ell^2}&=\frac{c_\ell}{\ell_s^2}-\frac{1}{\ell_u^2}, \qquad \frac{c_\ell-1}{\ell_u^2}=\frac{c_{\mathrm{cc}}}{\ell^2}-\frac{c_\ell}{3} \Lambda,\\
\frac{\Lambda}{3}&=\frac{1}{\ell_s^2}-\frac{1}{\ell_u^2},\qquad \frac{c_\ell-1}{\ell_s^2}=\frac{c_{\mathrm{cc}}}{\ell^2}-\frac{\Lambda}{3}.
\end{aligned}
\end{equation}
The solutions of Ref.~\cite{Max_Sym_UH} are classified according to the sign of the effective cosmological constant $\Lambda$. The asymptotically AdS case corresponds to $\Lambda<0$, for which $\ell_u < \ell_s$, and $\ell_s$ remains a free asymptotic parameter. The globally maximally symmetric solution in this branch is
\begin{equation}\label{eq:global ads}
(u\cdot\chi)=-\sqrt{\frac{r^2}{\ell_u^2}+1}, \qquad (s\cdot\chi)=\frac{r}{\ell_s}.
\end{equation}
Using Eq.~\eqref{eq:l relations}, one may also write
\begin{equation}
\ell_u^2 = \frac{\ell^2 \ell_s^2}{c_\ell \ell^2-c_{\mathrm{cc}} \ell_s^2}.
\end{equation}

An exact black-hole solution with a universal horizon in asymptotically AdS spacetime is given by \cite{Max_Sym_UH}
\begin{equation}\label{eq:ads UH soln}
\begin{aligned}
&(u\cdot\chi)=-\frac{r}{\ell_u}\left(1-\frac{r_{\textsc{uh}}}{r}\right)\sqrt{1+\frac{2r_{\textsc{uh}}}{r}+\frac{3r_{\textsc{uh}}^2+\ell_u^2}{r^2}+\frac{2r_{\textsc{uh}}(3r_{\textsc{uh}}^2+\ell_u^2)}{3r^3}+\frac{r_{\textsc{uh}}^2(3r_{\textsc{uh}}^2+\ell_u^2)}{3r^4}}\\
&(s\cdot\chi)=\frac{r}{\ell_s}+\frac{r_{\textsc{uh}}^2}{r^2\sqrt{3(1-c_{13})}}\sqrt{1+\frac{3r_{\textsc{uh}}^2}{\ell_u^2}}\\
&e(r)=-\frac{\Lambda r^2}{3}+1-\frac{r_0}{r}-\frac{c_{13}}{3(1-c_{13})}\left(1+\frac{3r_{\textsc{uh}}^2}{\ell_u^2}\right)\frac{r_{\textsc{uh}}^4}{r^4}\\
&F(r)=1\\
&r_0=\frac{4r_{\textsc{uh}}}{3}+\frac{2r_{\mathrm{\textsc{uh}}}^3}{\ell_u^2}+\frac{2r_{\textsc{uh}}^2}{\ell_s\sqrt{3(1-c_{13})}}\sqrt{1+\frac{3r_{\textsc{uh}}^2}{\ell_u^2}}.
\end{aligned}
\end{equation}
The parameter $r_{\textsc{uh}}$ denotes the location of the universal horizon. Indeed, Eq.~\eqref{eq:ads UH soln} gives
\begin{equation}
(u\cdot\chi)\big|_{r=r_{\textsc{uh}}}=0,
\end{equation}
as required. In the limit $r_{\textsc{uh}}\to 0$, the universal horizon disappears and the globally maximally symmetric solution in Eq.~\eqref{eq:global ads} is recovered.

Given the relationship between Einstein--\ae ther theory and Ho\v{r}ava--Lifshitz gravity discussed in Sec.~\ref{subsec:Aether Intro}, it is natural that universal horizons occur in both theories. A further important, yet less obvious result of Ref.~\cite{Max_Sym_UH} is that, in static spherically symmetric spacetimes, the black-hole solution spaces of Einstein--\ae ther theory and Ho\v{r}ava--Lifshitz gravity are isomorphic. Therefore, the solution above describes static, spherically symmetric, asymptotically AdS universal horizons in both Einstein--\ae ther theory and Ho\v{r}ava--Lifshitz gravity.

As a final remark, let us stress that, although the vacuum Einstein--\ae{ther} theory generically contains three propagating modes with finite characteristic speeds, and hence admits a corresponding nested set of Killing horizons, the asymptotically AdS solutions considered here belong to the $c_{14}=0$ sector. In this case, at least one mode, i.e., the scalar one, propagates with infinite speed. The universal horizon is therefore the relevant causal boundary for this class of solutions.

\subsection{First law for asymptotically AdS universal horizons}\label{subsec:UH thermo}
The thermodynamics of universal horizons is relatively well understood in asymptotically flat spacetimes~\cite{Mechanics,Max_Sym_UH,DelPorro:2023lbv}. By contrast, a coherent first law for universal horizons in asymptotically AdS spacetimes has been more difficult to formulate. A mass formula was originally proposed in Ref.~\cite{Max_Sym_UH}; varying this formula with respect to the horizon location led to a proposed first law. However, the result was fully understood only when the \ae ther aligned with the timelike Killing vector at infinity, corresponding to the limit $\ell_s\to\infty$. For a misaligned \ae ther at infinity, the variation of the mass formula contained an additional third term with no clear thermodynamic interpretation. As a result, the first law in the general asymptotically AdS case remained obscure. Moreover, the thermodynamic significance of asymptotic \ae ther alignment was itself unclear.

The mass formula for asymptotically AdS universal horizons was later adjusted in Ref.~\cite{Martin2024-tb} to incorporate the appropriate background subtraction. The corrected mass is
\begin{equation}\label{eq:ads uh mass}
M=\frac{1}{G_\textnormal{\ae}}\left[\frac{2r_{\textsc{uh}}}{3}+\frac{r_{\textsc{uh}}^3}{\ell_u^2}+\frac{r_{\textsc{uh}}^2}{\ell_s\sqrt{3(1-c_{13})}}\sqrt{1+\frac{3r_{\textsc{uh}}^2}{\ell_u^2}}\right].
\end{equation}
Varying this expression with respect to the universal horizon radius gives
\begin{equation}\label{eq:first law ads}
\delta M=\frac{q_{\textsc{uh}}\,\delta A}{8\pi G_\textnormal{\ae}}\,,
\end{equation}
where
\begin{equation}\label{eq:quh}
q_{\textsc{uh}}=\frac{2}{3r_{\textsc{uh}}}+\frac{3r_{\textsc{uh}}}{\ell_u^2}+\frac{2}{\ell_s\sqrt{3(1-c_{13})}}\left[1+\frac{9r_{\textsc{uh}}^2}{\ell_u^2}\right]\left[1+\frac{3r_{\textsc{uh}}^2}{\ell_u^2}\right]^{-1}.
\end{equation}
The first term above yields an entropy in the first law (although there are subtleties as discussed below)
while the second is the expected enthalpy contribution in asymptotically AdS spacetimes, where the cosmological constant plays the role of spacetime ``pressure''~\cite{Traschen}. The third term, however, had no clear thermodynamic interpretation. When the \ae ther aligns with the timelike Killing vector at infinity, i.e., $\ell_s\to\infty$, this term vanishes and one obtains a first law with a manifest thermodynamic interpretation.  Then the first law takes the form ~\cite{Max_Sym_UH,Martin2024-tb} 
\begin{equation}\label{eq:first law form aligned}
\delta M=\frac{\kappa}{8\pi G_\textnormal{\ae}}\delta A+\frac{\mathcal{V} }{8\pi G_\textnormal{\ae}}\delta\Lambda,
\end{equation}
where we reinstated the possibility of a non-fixed cosmological constant. 

Before considering the implications of misalignment, a comment is in order
about the subtleties in the entropy term in the first law. The surface gravity
$\kappa$ in the expression above is not the one associated with the peeling of
infinite-speed signals at the universal horizon,
$\kappa_{\textsc{uh}}=
\left.u^a\nabla_a (u\cdot\chi)\right|_{\textsc{uh}} .
$
Consequently, it is not associated with any physical temperature, and, in particular, it is not directly interpretable as the Hawking temperature of the universal horizon~\cite{DelPorro:2023lbv}. Indeed, it was shown in Ref.~\cite{Cropp:2013zxi} that the surface gravity
appearing in Eq.~\eqref{eq:first law form aligned}, as defined in
Ref.~\cite{Mechanics}, coincides with the \emph{metric} notion of surface gravity
evaluated at the universal horizon rather than at the Killing horizon. Namely,
for the solutions considered here one has $\kappa=e'(r_{\textsc{uh}})/2$.

The thermodynamic interpretation of this result was recently clarified in
Ref.~\cite{Arata:2026aza}. There, it was shown that the entropy contribution in
Eq.~\eqref{eq:first law form aligned} can be decomposed into two terms: a
genuine entropy term characterized by $\kappa_{\textsc{uh}}$, and a second
contribution associated with the Killing flux of the \ae ther field through the
universal horizon. By means of the Clausius relation, the latter can also be
recast as an entropy term proportional to $\kappa_{\textsc{uh}}$. The first law
can therefore be written as
\begin{equation}
\label{eq:first law form aligned-true}
\delta M
=
T_{\textsc{uh}}\delta A_{\mathrm{\textsc{uh}}}
+
T_{\textsc{uh}}\delta S_{\textnormal{\ae}}
+
\frac{\mathcal{V}}{8\pi G_{\textnormal{\ae}}}\delta\Lambda ,
\end{equation}
where $T_{\textsc{uh}}=\kappa_{\textsc{uh}}/\pi$~\cite{DelPorro:2023lbv}, while
$S_{\textnormal{\ae}}$ denotes the entropy contribution associated with the
Killing current of the \ae ther through the universal horizon. See
Ref.~\cite{Arata:2026aza} for further details. 

Independently of how one decomposes the $\kappa\,\delta A$ term into entropy contributions and isolates the \ae ther contribution, the third term in Eq.~\eqref{eq:quh} remains unexplained in Einstein--\ae ther theory. It is this term we focus on in this work.

\subsection{The misaligned case}

In Ho\v{r}ava--Lifshitz gravity, the parameter $\ell_s$ controls not only the asymptotic alignment of the \ae ther, but also the conservation of the \textit{elliptic charge} associated with the khronon reparameterization symmetry and its elliptic equation of motion. When this charge is conserved, no additional contribution to the first law is expected. This expectation was confirmed in Ref.~\cite{Martin2024-tb}, thereby identifying the extra term in Eq.~\eqref{eq:first law} with the elliptic charge. It also clarifies the thermodynamic role, in Ho\v{r}ava--Lifshitz gravity, of aligning the \ae ther with the timelike Killing vector at infinity.

Ideally, the Komar mass should be expressed in terms of the elliptic charge $Q_f$, rather than the parameter $\ell_s$, with a complete first law allowing variations of $Q_f$. In Ho\v{r}ava--Lifshitz gravity, asymptotic alignment corresponds to conservation of this charge, i.e., $Q_f$ remains constant. Imposing alignment thus amounts to working in a thermodynamic ensemble in which $Q_f$ is held fixed. In that ensemble the first law is well behaved.

Although the thermodynamics of asymptotically AdS universal horizons is now better understood in Ho\v{r}ava--Lifshitz gravity, its interpretation in Einstein--\ae ther theory has remained unclear. Universal horizons also arise in Einstein--\ae ther theory and, in the static spherically symmetric asymptotically AdS sector, obey the same first-law, Eq.~\eqref{eq:first law}, since the black-hole solution spaces of the two theories are isomorphic \cite{Max_Sym_UH}. However, Einstein--\ae ther theory lacks the khronon reparameterization symmetry of Ho\v{r}ava--Lifshitz gravity, and hence has no associated elliptic charge. It has therefore been unclear why a coherent first law emerges in the limit $\ell_s\to\infty$.

The resolution, developed in the following sections, is that for $c_{14}=0$ and in spherical symmetry Einstein--\ae ther theory acquires an additional \ae ther symmetry with an associated conserved charge. Setting $c_{14}=0$ therefore introduces a charge whose contribution at the spatial boundary vanishes when the \ae ther is asymptotically aligned with the timelike Killing vector. This charge provides the Einstein--\ae ther analog of the thermodynamic mechanism that, in Ho\v{r}ava--Lifshitz gravity, is tied to conservation of the elliptic charge.

As we shall see, in Einstein--\ae ther theory the alignment condition is even stronger: it sets $Q_f=0$. One may therefore restrict attention to the subspace of solutions with fixed charge, here the uncharged sector $Q_f=0$. This is analogous to treating Schwarzschild black holes as the zero-charge subsector of the Reissner--Nordstr\"om family, with the corresponding first law obtained by restricting to fixed electric charge.

\section{\AE ther symmetry in spherically symmetric asymptotically AdS spacetimes}\label{sec:symmetry}

In this section we show that the parameter $\ell_s$ is not merely an artifact of the solution-generating ansatz, but is instead associated with a genuine symmetry in the reduced (spherically symmetric and asymptotically AdS) sector.

To do so, we first recall the covariant phase space construction of Noether currents in the presence of boundaries, emphasizing the role of boundary and corner terms in defining an unambiguous current.  We then give a pedagogical exposition for the existence of the symmetry by rewriting the reduced Lagrangian and neglecting boundary terms. With the symmetry in hand, we then compute the boundary and corner terms, Noether current, and flux in the full covariant phase space formalism.  Finally, we reinterpret the same structure in Hamiltonian language, where the charge appears as a weakly vanishing constraint generator whose Hamiltonian flow reproduces the \ae ther transformation.

\subsection{Symmetries and Noether currents}

We begin by recalling the covariant phase space construction of Noether
currents. For a theory defined by a Lagrangian top form on $(d+1)$-dimensional spacetime $\mathcal{M}$, one obtains the equations of motion of a theory by varying its Lagrangian:
\begin{equation}\label{eq:lagrangian variation}
\delta L = E\cdot \delta\phi +\mathrm{d}\Theta
\end{equation}
where $E=0$ denotes the equations of motion, $\phi$ collectively denotes the dynamical fields of the theory, and the dot ``$\cdot$'' suppresses the sum over field-space indices. In the present case, $\phi$ consists of the metric $g_{ab}$ and the \ae ther $u_a$. The $d$-form $\Theta$ is the symplectic potential. Here the variation $\delta$ is best regarded as an exterior derivative on the configuration space.

\begin{figure}[H]
    \centering
    \begin{tikzpicture}[scale=0.6]

        \draw (0,0) arc(0:180:4cm and 0.5cm);
        \draw (-8,0) arc(180:360:4cm and 0.5cm);

        \draw (0,0) .. controls (-1,-6) .. (0,-12);
        \draw (-8,0) .. controls (-7,-6) .. (-8,-12);

         \draw[dashed] (0,-12) arc(0:180:4cm and 0.5cm);
         \draw (-8,-12) arc(180:360:4cm and 0.5cm);

         \draw[dashed] (-0.75,-6) arc(0:180:3.25cm and 0.5cm);
         \draw (-7.25,-6) arc(180:360:3.25cm and 0.5cm);
        
        \draw (0.5, -2) node {{$\Gamma$}};
        \draw (-4, -12) node {{$\Sigma_-$}};
        \draw (-4, 0) node {{$\Sigma_+$}};
        \draw (-4, -6) node {{$\Sigma$}};
        \draw (0.7, 0) node {{$\partial\Sigma_+$}};
        \draw (0.7,-12) node {{$\partial\Sigma_-$}};
        
        \draw[->] (-8.5,-12) -- (-8.5,0);
        \draw (-9, 0) node {{$t$}};

        \draw[->] (-4,0.4) -- (-4,1.05) node [anchor = east] {{$t$}};
        \draw[->] (-4,-11.6) -- (-4,-10.95) node [anchor = east] {{$t$}};
        \draw[->] (-4,-5.65) -- (-4,-5) node [anchor = east] {{$t$}};
        \draw[->] (-0.75,-6) -- (-0.25,-6) node [anchor = west] {{$n$}};
    \end{tikzpicture}
    \caption{Representation of the cylindrical spacetime under consideration, with its different parts: the lateral boundary $\Gamma$, the two lids $\Sigma_\pm$, the two corners $\partial\Sigma_i$, and a generic slice $\Sigma$.}
    \label{fig: spacetime}
\end{figure}
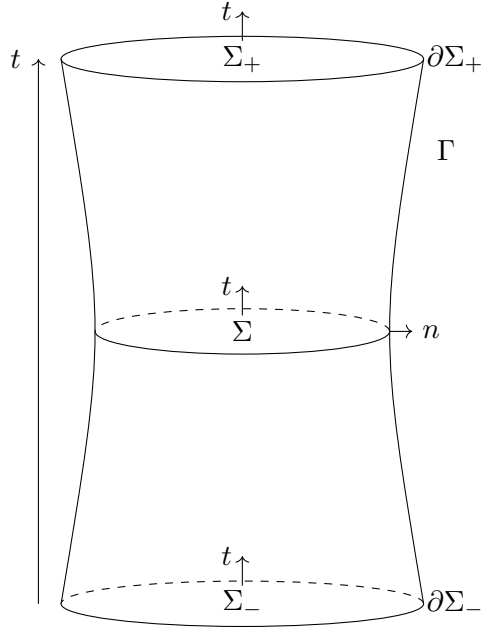

The equations of motion follow by requiring the action to be stationary up to boundary terms on the asymptotic past and future. To make the variational principle well-defined, one generally supplements the bulk Lagrangian $L$ by a boundary contribution $L_\partial$, so that the action is 
\begin{equation}\label{eq:general action}
S=\int_{\mathcal{M}}L+\int_{\Gamma} L_{\partial},
\end{equation}
where $\Gamma\subset\partial\mathcal{M}$ denotes the subset of $\partial\mathcal{M}$ which is either the timelike boundary or a future null boundary (see Fig. \ref{fig: spacetime}). In the asymptotically AdS spacetimes considered here, $\Gamma$ is the timelike boundary. The boundary Lagrangian $L_{\partial}$ is chosen so that the variational principle is well-defined and compatible with the desired boundary conditions. A familiar example is the Gibbons--Hawking--York (GHY) term in general relativity \cite{Gibbons_Hawking,York}.

On $\Gamma$, the symplectic potential can in general be decomposed as \cite{Tony,Harlow_Covariant}
\begin{equation}\label{eq:symp pot restriction}
\Theta|_{\Gamma} = -\delta L_{\partial} +\mathcal{E} +\mathrm{d}\theta
\end{equation}
where $\mathcal{E}$ is the flux term and $\theta$ is the corner symplectic potential. Using Eqs.~\eqref{eq:lagrangian variation},~\eqref{eq:symp pot restriction}, and Stokes' theorem, one obtains~\cite{Tony, Harlow_Covariant}
\begin{equation}\label{eq:action variation}
\delta S=\int_{\mathcal{M}}E\cdot\delta\phi+\int_\Gamma\mathcal{E}+\int_{\Sigma_-}^{\Sigma_+} (\Theta - \mathrm{d}\theta).
\end{equation}
Here $\Sigma_\pm$ are the future and past lids which, together with $\Gamma$, form the boundary of $\mathcal{M}$. They are the hypersurfaces on which one specifies the appropriate initial data to choose a state in the theory. Thus, the equations of motion hold when the flux term through $\Gamma$ vanishes under the imposed boundary conditions, while the term $-\delta L_{\partial}$ cancels the boundary variations incompatible with those conditions~\cite{Tony,Harlow_Covariant}.

A theory is said to possess a symmetry if there exists a vector field $\hat{\zeta}$ on configuration space such that~\cite{Tony,Harlow_Covariant}
\begin{equation}\label{eq:Symmetry defn}
I_{\hat{\zeta}}\delta L = \mathrm{d}K_{\hat{\zeta}}
\end{equation}
for some codimension-one spacetime form $K_{\hat{\zeta}}$. Here $I_{\hat{\zeta}}$ denotes the interior product of configuration space forms and the vector field $\hat{\zeta}$. In other words, the transformation of the dynamical fields generated by $\hat{\zeta}$ changes the Lagrangian by an exact spacetime form $\mathrm{d}K_{\hat{\zeta}}\in\Omega^{d+1}(\mathcal{M})$. On the other hand, contracting Eq.~\eqref{eq:lagrangian variation} with $\hat{\zeta}$ gives
\begin{equation}\label{eq:dL with EOM}
I_{\hat{\zeta}} \delta L = E \cdot I_{\hat{\zeta}} \delta \phi +\mathrm{d}I_{\hat{\zeta}}\Theta
\end{equation}
Combining Eqs.~\eqref{eq:Symmetry defn} and~\eqref{eq:dL with EOM}, it is natural to define the Noether current $d$-form associated with $\hat{\zeta}$ by
\begin{equation}\label{eq:J defn}
J_{\hat{\zeta}}=I_{\hat{\zeta}}\Theta-K_{\hat{\zeta}}.
\end{equation}
Its exterior derivative is
\begin{equation}\label{eq:J closed}
\mathrm{d}J_{\hat{\zeta}}=\mathrm{d}I_{\hat{\zeta}}\Theta-\mathrm{d}K_{\hat{\zeta}}=-E\cdot I_{\hat{\zeta}}\delta\phi.
\end{equation}
Thus $J_{\hat{\zeta}}$ is closed on shell, i.e., $\mathrm{d}J_{\hat{\zeta}}=0$ when $E=0$. This is the covariant phase space version of the familiar continuity equation.

\subsection{Identifying the symmetry}
The covariant phase space approach does not, of course, allow one to determine what vector fields on configuration space generate symmetries.  In order to find the previously unidentified symmetry, we first rewrite the {\ae}ther Lagrangian, Eq.~\eqref{eq:Lae twist etc}, using
the Raychaudhuri equation for a unit timelike congruence $u^a$, 
\begin{align}
    u^a \nabla_a \vartheta = -\frac{1}{3}\vartheta^2 -\sigma_{ab}\sigma^{ab} +\omega_{ab}\omega^{ab} -R_{ab} u^a u^b +\nabla_a a^a.
\end{align}
Solving for $\sigma_{ab}\sigma^{ab}$, the kinetic part of the \ae ther
Lagrangian may be written as
\begin{align}
    \mathcal{L}_{\textnormal{\ae}}= \left(- c_\vartheta +\frac{c_\sigma}{3}\right) \vartheta^2 &+c_\sigma \left(u^a \nabla_a \vartheta +R_{ab}u^a u^b -\nabla_a a^a\right) - \left(c_\omega +c_\sigma\right) \omega_{ab}\omega^{ab} + c_a a^2 
\end{align}

We now restrict to static, spherically symmetric configurations. In this
sector the twist vanishes identically, $\omega_{ab}=0$. We also specialize to the $c_a = c_{14} = 0$ sector which corresponds to that of the black-hole solutions considered in this paper.\footnote{Recall from Section~\ref{subsec:UH} that solutions for universal horizons exist in asymptotically AdS spacetimes only in the case where $c_{14}=0$ or equivalently $c_a = 0$.} The kinetic part of the Lagrangian becomes
\begin{align}\label{eq: reduced AEther Lagrangian}
    \mathcal{L}_{\textnormal{\ae}}= \left(- c_\vartheta +\frac{c_\sigma}{3}\right) \vartheta^2 +c_\sigma \left(u^a \nabla_a \vartheta +R_{ab}u^a u^b -\nabla_a a^a\right).
\end{align}
Using the Leibniz rule, $u^a \nabla_a \vartheta =  \nabla_a (\vartheta u^a) -\vartheta^2 $, we can split the Lagrangian into a bulk term and a total derivative:
\begin{align}\label{eq: Lagrangian decomposition}
    \mathcal{L}_{\mathrm{\textnormal{\ae}}} = \mathcal{L}_0 + \nabla_a A^a,
\end{align}
where, adding the Lagrange multiplier back in,
\begin{align}\label{L0}
    \mathcal{L}_0 \equiv  -\left(c_\vartheta +\frac{2}{3} c_\sigma\right) \vartheta^2 +c_\sigma R_{ab}u^a u^b +\lambda_{\textnormal{\ae}}(u^2+1), \qquad A^a \equiv c_\sigma (\vartheta u^a -a^a).
\end{align}
Variations of the total derivative terms will not matter for finding the symmetry (although they will come into play in defining the current and charge), and so we can temporarily ignore them. 

The latter two terms in $
\mathcal{L}_0$ are manifestly invariant under transformations of the form %
\begin{equation}\label{eq: u xfm}
u_a\mapsto u_a+fa_a, \quad
g_{ab} \mapsto g_{ab}
\end{equation}
where $f$ is a scalar function. To show that, let us denote these transformations by $\delta_f$. Since the acceleration is orthogonal to
the {\ae}ther, the unit-norm constraint is preserved at first order:
\begin{equation}\label{eq:unit norm pres}
\delta_f(u^a u_a) = 2 u_a \,\delta_f u^a = 2 f \, u_a a^a = 0\,.
\end{equation}
Additionally, for any (possibly time-dependent) spherically symmetric geometry, a purely kinematical result is that the mixed temporal-radial projection of the Ricci tensor
vanishes~\cite{Abreu_2010}.  Thus, $R_{ab} u^a \delta_f u^b =f  R_{ab} u^a a^b =0$. Hence, any relevant dependence on
$\nabla_a u_b$ in $\mathcal{L}_0$ appears only through the expansion
$\vartheta=\nabla_a u^a$. Therefore,
\begin{align}
    \delta_f \mathcal{L}_{\textnormal{\ae}}= -2\left(c_\vartheta +\frac{2}{3} c_\sigma\right) \vartheta\, \delta_f\vartheta + \mathrm{total~derivative}.
\end{align}
Consequently, the transformation is a symmetry provided that $\delta_f\vartheta=0$. Since the metric is fixed, this condition directly implies the elliptic equation\footnote{To see that~Eq.~\eqref{elliptic} is in fact elliptic, we can expand $\nabla_a(fa^a)=g^{ab}\nabla_a(fa_b)=(h^{ab}-u^au^b)\nabla_a(fa_b)$. The first term becomes an entirely spatial derivative on $\Sigma$, $\vec{\nabla}_a(fa^a)$. Then, using the chain rule and thefact that $(a\cdot u)=0$, the second term becomes $u^au^b\nabla_a a_b=-a^2$. We can therefore recast Eq.~\eqref{elliptic} in the more obviously elliptic form:
\[
\vec{\nabla}_a(fa^a)+fa^2=0.
\]}
\begin{equation}\label{elliptic}
    \delta_f\vartheta = \nabla_a(fa^a)=0.
\end{equation}
When this equation holds, the variation of the Lagrangian is simply that of the total
derivative:
\begin{align}\label{eq: variationTotalDerivative}
    \delta_f \mathcal{L}_{\textnormal{\ae}} = \nabla_a \delta_f A^a, \qquad \delta_f A^a \equiv c_\sigma (\vartheta \delta_f u^a -\delta_f a^a).
\end{align}
Thus, the transformation $\delta_f u_a=fa_a$, supplemented by
Eq.~\eqref{elliptic}, is a symmetry of the reduced \ae ther Lagrangian, Eq.~\eqref{eq: reduced AEther Lagrangian}.

After determining the symmetry of this theory sector, we  can compute the associated current and charge. To do so, we will use the procedure introduced in the previous subsection, specifically Eq.~\eqref{eq:J defn}, using the decomposition Eq.~\eqref{eq: Lagrangian decomposition}.

Let us start from the fact that $\delta_f \mathcal{L}_0 = 0$, as previously imposed via the elliptic constraint, Eq.~\eqref{elliptic}. Using this in  Eq.~\eqref{eq: variationTotalDerivative} we can immediately read from Eq.~\eqref{eq:Symmetry defn} that
\begin{equation}
    K^a_f = \delta_f A^a\,.
\end{equation}
The symplectic potential $\Theta$ is the total derivative one gets after performing all the possible integrations by parts that act only on the field variations $\delta u^a$ when varying the Lagrangian. In particular, 
\begin{equation}
    \dfrac{\delta(\sqrt{-g} \, \mathcal{L}_0)}{\sqrt{-g}} = \nabla_a \underbrace{\left[- 2 \, \left(c_\vartheta + \dfrac{2}{3} \, c_\sigma\right) \, \vartheta \, \delta u^a + X_0^{abc} \, \delta g_{bc}\right]}_{\Theta_0^a}  + \left(\textnormal{\AE}_a + \lambda_\textnormal{\ae} \, u_a \right) \, \delta u^a + \mathcal{T}^{ab} \, \delta g_{ab},
\end{equation}
where $\mathcal{T}^{ab}$ is the stress-energy tensor of the \ae ther.
The exact form of the tensor $X_0^{abc}$ is not required because we are interested in contracting this variation with $I_f$, which would set $\delta g_{bc}=0$. The total symplectic potential is therefore
\begin{equation}
    \Theta^a = \Theta^a_0 + \delta A^a\,.
\end{equation}
Hence, the current becomes
\begin{equation}\label{eq: f-current}
    J^a_f = I_f \Theta^a - K^a_f = I_f\Theta^a_0 + \cancel{\delta_f A^a} - \cancel{\delta_f A^a} = - 2 \, \left(c_\vartheta + \dfrac{2}{3} \, c_\sigma \right) \, \vartheta \, \delta_f u^a = - 2 \, \left(c_\vartheta + \dfrac{2}{3} \, c_\sigma \right) \, \vartheta \, f \, a^a.
\end{equation}

The final form of the current, after all the simplification, is then
\begin{equation}
    J^a_f = - 2 \, \left(c_\vartheta + \dfrac{2}{3} \, c_\sigma \right) \, \vartheta \, f \, a^a = - 2 \, c_{123} \, \vartheta \, f \, a^a,
\end{equation}
where the coefficient $c_{123}$ appeared using the relations Eq.~\eqref{eq:couplings dictionary}. Note that in the $c_{123}=0$ sector this current is identically zero, making the transformation unphysical. This result is consistent with the fact that when $c_{14} = c_{123} = 0$ the theory is equivalent to general relativity under a field redefinition in the twist-free sector; cf. Sec.~VI of Ref.~\cite{Sotiriou:2014gna}.

The associated Noether charge follows by integrating the current over a spacelike hypersurface $\Sigma$,
\begin{equation}
  Q_{f}=\int_{\Sigma}n_{a}\,J^{a}_{f}\,\mathrm{vol}_{\Sigma}.
  \label{eq:charge}
\end{equation}
In the \ae ther frame, which exists globally because $\omega_{ab}=0$, one may take $n_{a}=u_{a}$; the current is then orthogonal to the normal, $u_{a}J^{a}_{f}\propto u_{a}a^{a}=0$, so the charge vanishes on \ae ther-orthogonal slices,
\begin{equation}
  Q_{f}=0.
  \label{eq:Qzero}
\end{equation}
This does not render the current trivial. It tells us that the current is purely spatial in the \ae ther frame, and that the relevant quantity is its flux through a timelike surface $\Gamma$. If we take the latter to be a surface with unit normal $s^{a}$, then we get
\begin{equation}
  \mathcal{F}_{\textnormal{\ae}} = \int_{\Gamma}s_{a}\,J^{a}_{f} \, \mathrm{vol}_{\Gamma} = - 8 \pi \, c_{123} \, (u \cdot \chi) \, r^{2}  \, f \, \vartheta \, (a\cdot s) \,,
  \label{eq:flux}
\end{equation}
which is independent of $r$ on-shell. The $(u \cdot \chi) \, r^2$ factor comes from the volume element $\mathrm{vol}_\Gamma$ written in the \ae ther frame (see Eq.~(4.19) in Ref.~\cite{Arata:2026aza}). 

\subsection{Hamiltonian formalism, elliptic equations of motion, and symmetry generators}

While $Q_f=0$ in the \ae ther frame, it still generates the flow in the symmetry direction. To see this most clearly, we apply the Hamiltonian formalism where the relation between the \ae ther charge and the proposed symmetry can be stated more explicitly. We work in units where $16\pi G_{\textnormal{\ae}}=1$, and we take the components of the \ae ther $u_a$ to be the generalized coordinates. To define the canonical momentum conjugate to $u_a$, one must first choose a timelike vector field along which evolution is generated. Ultimately, we will specialize to the preferred frame, in which the evolution vector is the \ae ther itself and the \ae ther is hypersurface-orthogonal to spacelike slices $\Sigma$. This choice is natural in the present context, both for convenience and because the Hamiltonian structure away from the preferred foliation is not well understood. For the moment, however, we keep the evolution vector and the slicing general.

Let $\Sigma$ be a spacelike hypersurface and let $v^a$ denote a timelike vector field generating evolution relative to this hypersurface. The momentum density conjugate to the \ae ther is obtained by differentiating the Lagrangian with respect to $\nabla_a u_b$. In the spherically symmetric sector with $c_a=c_{14}=0$, it simplifies to
\begin{equation}\label{eq:momentum density}
\Pi^{ab} \equiv \frac{\partial \mathcal{L}_0}{\partial(\nabla_a u_b)}
=
- 2 \, \left(c_\vartheta + \dfrac{2}{3} c_\sigma\right)\, \vartheta \, h^{ab}\,.
\end{equation}
The canonical momentum associated with evolution along $v^a$ is therefore $\pi^a\equiv v_b\Pi^{ba}$.

In a general frame, the canonical Hamiltonian is obtained by performing the Legendre transformation on a spacelike slice $\Sigma$: 
\begin{equation}\label{eq:H ae general frame}
H_{\textnormal{\ae}}=\int_{\Sigma} \vol_{\Sigma}\left(\pi^av^b\nabla_b u_a-\mathcal{L}_0\right)+\underline{H}_{\textnormal{\ae}}\,.
\end{equation}
The term $\underline{H}_{\textnormal{\ae}}$ denotes a possible boundary Hamiltonian. Such a term is required whenever boundary contributions are needed to make the variational principle and the canonical momenta well defined. In the gravitational sector, the analogous contribution arises from the Gibbons--Hawking--York term and is related to the total energy of the spacetime. In the present discussion, we do not need to impose a specific choice of \ae ther boundary conditions, and so we leave $\underline{H}_{\textnormal{\ae}}$ unspecified. Notice that the boundary term given by $\nabla_a A^a$ has been included in $\underline{H}_{\textnormal{\ae}}$.

We now specialize the evolution vector to be the \ae ther itself, $v^a = u^a$, and, in the twist-free case, denote the corresponding hypersurface by $\Sigma_u$.
In this frame, the canonical momentum becomes
\begin{equation}
    \pi^a = u_b \Pi^{ba}.
\end{equation}
Since $h^{ab}$ is completely spatial by construction, the canonical momentum is orthogonal to $u^a$. 
Hence, we get the primary constraints,
\begin{equation}
    C_0^a \equiv \pi^a \approx 0,
\end{equation}
where we introduced the Dirac weak equality symbol ``$\approx$'' to denote equality on the constraint surface. The canonical Hamiltonian therefore takes the form
\begin{equation}\label{eq:H ae}
H_{\textnormal{\ae}}=\int_{\Sigma_u} \vol_\Sigma\left(\tilde{\lambda}_a \pi^a-\mathcal{L}_0\right)+\underline{H}_{\textnormal{\ae}} \approx -\int_{\Sigma_u}\vol_{\Sigma_u} \mathcal{L}_0+\underline{H}_{\textnormal{\ae}}
\end{equation}
where $\tilde{\lambda}_a$ are Lagrange multipliers enforcing the primary constraints. The second equality holds on the constraint surface.

Time evolution is generated by the Hamiltonian through Poisson brackets. Therefore, consistency of the constraint algebra requires that the primary constraints be preserved under time evolution. Applying Dirac's algorithm, we demand
\begin{equation}
    \dot C_0^a
    =
    \{C_0^a,H_{\textnormal{\ae}}\}
    \approx 0.
\end{equation}
Using the Hamiltonian in Eq.~\eqref{eq:H ae}, and suppressing boundary terms canceled by the appropriately chosen $\underline{H}_{\textnormal{\ae}}$, we obtain
\begin{equation}\label{eq:elliptic eqn u}
\begin{aligned}
\{C_0^a,H_{\textnormal{\ae}}\}&=-\Big\{\pi^a,\int_{\Sigma_u}\mathcal{L}_0 + \underline{H}_\textnormal{\ae}\Big\}\\
&=\int_{\Sigma_u}\frac{\delta\pi^a}{\delta\pi^b}\frac{\delta\mathcal{L}_0}{\delta u_b} + \{\pi^a, \underline{H}_\textnormal{\ae}\}\\
&=-\int_{\Sigma_u}\delta^a_{\;b}\left(-\nabla_c\Pi^{cd}+2 \, \lambda_{\textnormal{\ae}}u^d\right)\delta^b_{\;d}\,\delta(x-y)\\
&=\nabla_b\Pi^{ba}-2\lambda_{\textnormal{\ae}}u^a \,,
\end{aligned}
\end{equation}
which requires the imposition of a secondary constraint,
\begin{equation}\label{eq:C1}
C_1^a
=
\nabla_b\Pi^{ba}-2\lambda_{\textnormal{\ae}}u^a
\approx
0.
\end{equation}
This is equivalent to the \ae ther equation of motion $\text{\AE}_a+\lambda_\textnormal{\ae}u_a=0$ derived in Eq.~\eqref{eq:eoms}. Since these are the Euler--Lagrange equations of motion, it is clear that there are no other independent constraints to impose.

In the $c_a=0$ sector, Eq.~\eqref{eq:C1} contains an elliptic component. Contracting with $u_a$ simply allows one to solve for the Lagrange multiplier. However, contracting with $s^a$, which amounts to calculating $s_a \text{\AE}^a$ from Eq.~\eqref{eq:eom specifics-Aether}, yields the expression
\begin{equation}\label{eq: s nabla theta}
c_{123} \, s^a \, \nabla_a \vartheta \approx 0.
\end{equation}
This makes manifest in the Hamiltonian framework why the current in Eq.~\eqref{eq: f-current} is conserved given the elliptic equation condition on $f$.

Now that the Hamiltonian formulation is in place, we can show directly that the
\ae ther charge $Q_f$ generates the symmetry transformation, Eq.~\eqref{eq: u xfm}. We simply take the charge, as given by Eq.~\eqref{eq:charge}, reduce its integrand in the \ae ther frame, and compute its Poisson bracket with $u_a$.

Using the canonical momentum $\pi^a=u_b\Pi^{ba}$ and the transformation $\delta_f u_b=f a_b$, Eq.~\eqref{eq:charge} reduces to
\begin{equation}\label{eq:Q canonical}
Q_f
=\int_{\Sigma_u}\vol_{\Sigma_u}\,u_a\,J^a_f
=\int_{\Sigma_u}\vol_{\Sigma_u}\,u_a\Pi^{ab}\,f a_b
=\int_{\Sigma_u}\vol_{\Sigma_u}\,\pi^a\,f a_a .
\end{equation}
The integrand is exactly the canonical pairing $\pi^a\,\delta_f u_a$ of the
momentum with the symmetry transformation. In the \ae ther frame $Q_f\approx0$ because
$\pi^a\approx0$; as with the Gauss-law generator in electromagnetism, this weak
vanishing does not signal triviality, since a constraint generator can still act
non-trivially through its Hamiltonian vector field.

Bracketing $Q_f$ with the \ae ther yields
\begin{align}\label{eq:bracket computation}
\{u_a(x),Q_f\}
&=\int_{\Sigma_u}\vol_{\Sigma_u}(y)\,\Big[\{u_a(x),\pi^c(y)\}\,f(y)a_c(y)
+\pi^c(y)\,\{u_a(x),f(y)a_c(y)\}\Big]\notag\\
&=f(x)\,a_a(x)
+\int_{\Sigma_u}\vol_{\Sigma_u}(y)\,\pi^c(y)\,\{u_a(x),f(y)a_c(y)\}\notag\\
&\approx f(x)\,a_a(x),
\end{align}
where we dropped the second term because it is proportional to the momentum $\pi^c$ and thus vanishes on the constraint surface. 

This remains true despite the acceleration
$a_c=u^b\nabla_b u_c$ being velocity-like, so that $\{u_a,a_c\}$ need not vanish:
working momentarily at $c_a\neq0$, where the acceleration-squared term renders
the Legendre map non-degenerate in this sector and $a_c=a_c(u,\pi)$, any such
contribution is manifestly weighted by $\pi^c$, and the limit $c_a\to0$ is
smooth. We therefore obtain
\begin{equation}\label{eq:Q generates}
\{u_a,Q_f\}\approx f\,a_a=\delta_f u_a.
\end{equation}
Thus, the \ae ther charge $Q_f$ generates the symmetry Eq.~\eqref{eq: u xfm} weakly,
even though it vanishes on the constraint surface. 

\section{\AE ther alignment as choice of ensemble} \label{sec:ensemble}

Now that we have established the existence of the charge $Q_f$, we apply it to black-hole thermodynamics.  Fundamentally, a first law of black-hole thermodynamics describes the smooth transition between regular black hole solutions in terms of smooth variations of the underlying degrees of freedom. To account for all solutions, the most general first law must describe how variations of \emph{all} degrees of freedom contribute to the variation of the total energy (mass) of the black hole. 

However, as is typical in thermodynamics, one can choose an ensemble that freezes certain degrees of freedom. This is also typical in black-hole thermodynamics; for instance, the $Q=\delta Q=0$ sector of the Reissner--Nordstr\"om black hole gives consistent thermodynamics for the Schwarzschild black hole in general relativity. Hence, while the existence of an \ae ther charge $Q_f$ in Einstein--\ae ther theory means that an ultimate first law of universal horizons must describe how variations of $Q_f$ contribute to variations of the universal horizon mass, the goal of this paper is simply to demonstrate that the choice of asymptotic \ae ther alignment with the Killing vector has a physical interpretation: it is the choice of the ensemble where $Q_f=\delta Q_f=0$.

To make the connection between asymptotic alignment and the \ae ther charge
precise, we must identify the invariant that carries the misalignment. In the
\ae ther frame, as Eq.~\eqref{eq:Qzero} shows, the bulk charge vanishes identically for
\emph{every} solution, aligned or not, because $J^a_f\propto a^a$ is orthogonal
to $u^a$. Vanishing of $Q_f$ in this frame is thus a frame artifact and
cannot by itself diagnose alignment. 

Let us then work in the Killing frame, although any frame would work equivalently.  If we define $\mathcal{R}^a$ as the spatial vector such that $\chi \cdot \mathcal{R}=0$, then we can construct the charge density $\rho_\chi$ and radial flux density $F_\chi$ via
\begin{equation}
\rho_\chi=-\chi_a J^a=-(s \cdot \chi)  J_s, \qquad F_\chi=\mathcal{R}_a J^a= (u \cdot \chi) J_s 
\end{equation}
where $J_s = s_a \, J^a = - 2 \, c_{123} \, f \, \vartheta \, (a \cdot s)$.
Since $(u \cdot \chi)$ vanishes nowhere outside the universal horizon, the only way to make the flux density vanish is to set $J_s=0$, which then sets the charge density to zero as well.

The misalignment enters into $J_s$ through the \ae ther expansion $\vartheta$. For static, spherically symmetric solutions, the
$s$-projection of the \ae ther equation of motion, Eq.~\eqref{eq: s nabla theta}, yields an equation for $(s \cdot \chi)$, 
\begin{equation}\left(\partial_r^2+\tfrac{2}{r}\partial_r-\tfrac{2}{r^2}\right) (s\cdot \chi) = 0\,,
\end{equation}
whose general solution is a linear combination of two independent solutions
\begin{equation}\label{eq:Sigma modes}
(s \cdot \chi) =\alpha\,r+ \frac{\beta}{r^2}.
\end{equation}
Given the asymptotic behavior of $(s\cdot\chi)$, one can check that the coefficient $\alpha$ defines the misalignment parameter by\footnote{See for instance the explicit solution, Eq.~\eqref{eq:ads UH soln}.}
\begin{equation}
    \alpha=\frac{1}{\ell_s}\,,
\end{equation}
The two modes, the growing and the decaying one, are potential contributions to the total elliptic charge and flux.

Notice that the decaying mode, with coefficient $\beta$, only occurs when there is a universal horizon, as it is irregular at $r=0$. It changes the near-horizon geometry, but not the geometry at asymptotic infinity.

Noticeably, the expansion is only sensitive to the growing mode,
\begin{equation}\label{eq:theta value}
\vartheta=-\Big[(s\cdot\chi)'+\tfrac{2}{r}(s\cdot\chi)\Big]
=-\frac{1}{r^2}\frac{\mathrm{d}}{\mathrm{d}r}\big[r^2(s\cdot\chi)\big]
=-3\alpha=-\frac{3}{\ell_s},
\end{equation}
because the decaying mode obeys $r^2(\beta r^{-2})=\beta$ and is thus
annihilated by the radial derivative.
Two consequences follow. First, the flux density is proportional to the misalignment,
\begin{equation}\label{eq:flux alignment}
F_\chi
\propto \vartheta (a\cdot s) f
\ \propto\ \vartheta=-\frac{3}{\ell_s} \,.
\end{equation}
Second, given that $(a\cdot s)$ is nowhere vanishing for black-hole solutions with a universal horizon~\cite{Bhattacharyya:2015gwa, Benkel:2018abt}, there is no way to preserve a vanishing flux when varying the asymptotic alignment by simultaneously adjusting horizon or bulk geometry. Hence, the relevant ensemble would be defined by one of the following equivalent conditions:
\begin{equation}\label{eq:flux iff}
\mathcal{F}=0
\quad\Longleftrightarrow\quad \vartheta=0
\quad\Longleftrightarrow\quad \ell_s\to\infty
\quad\Longleftrightarrow\quad \text{asymptotic alignment}.
\end{equation}
For example, in the solution given in Eq.~\eqref{eq:ads UH soln}, the explicit form of the solution $f$ to the elliptic equation, Eq.~\eqref{elliptic}, is
\begin{equation}
    f(r) = \dfrac{\mathcal{C} \, r^3}{(r^3 - r_{\textsc{uh}}^3)(3 \, r^3 + 2 \, \ell_u^2 \, r_{\textsc{uh}} + 6 \, r_{\textsc{uh}}^3)} \,,
\end{equation}
where $\mathcal{C}$ is an integration constant.
The flux turns out to be
\begin{equation}
    \mathcal{F} = \int_\Gamma \, r^2 \, F_\chi = \int_\Gamma \, (u \cdot \chi) \, r^2 \, f(r) \, \vartheta \, (a \cdot s) = \dfrac{\mathcal{C}}{\ell_s \, \ell_u^2}\,.
\end{equation}
Using the relation $\Lambda = 3/\ell_s^2 - 3/\ell_u^2$ in Eq.~\eqref{eq:l relations}, we can rewrite the flux as
\begin{equation}
    \mathcal{F} = \mathcal{C} \, \left(\alpha^3 - \dfrac{\Lambda}{3} \, \alpha \right) \,, 
\end{equation}
and its variation as
\begin{equation}
    \delta \, \mathcal{F} = \mathcal{C} \left( 3 \, \alpha^2 - \dfrac{\Lambda}{3} \, \right) \delta \alpha \,.
\end{equation}
If we consider the ensemble for which $\delta \alpha = 0$---more specifically the aligned ensemble, $\alpha = 0$---then both the flux and its variation are zero. Note that in the previously found asymptotically flat geometries the aether is both aligned, $\alpha=0$, and $\Lambda$ vanishes.  Therefore, the variation of the flux vanishes at first order, giving further evidence why the flux and charge associated with the symmetry $\delta_f$ were naturally absent in those solutions~\cite{Max_Sym_UH}.

Holding the \ae ther charge fixed at zero, which imposes zero flux, is then
identical to selecting the subset of aligned solutions, $\ell_s\to\infty$. This is independent of
any additional $\beta$-dependent terms or other variations that may appear in a full, dynamical first law.  In addition, maintaining zero flux sets the charge density and total charge to zero. Hence, asymptotic alignment selects the ensemble in which
$Q_f=\delta Q_f=0$. Finally, we note that this is the same behavior as the otherwise unexplained term in the first
law: the third term in $q_{\textsc{uh}}$ in Eq.~\eqref{eq:quh} is proportional to
$1/\ell_s=-\vartheta/3$ and hence to $\mathcal{F}_{\chi}$, switching
off precisely in the aligned limit.

\section{Summary and future directions}\label{sec:summary}

In this paper, we have taken a step toward formulating a coherent first law for universal horizons in asymptotically AdS spacetimes. Recent progress in Ho\v{r}ava--Lifshitz gravity showed that the previously unexplained term appearing in the variation of the universal horizon mass is tied to an elliptic charge associated with the khronon reparameterization symmetry of the theory. In particular, Ref.~\cite{Martin2024-tb} showed that conservation of this elliptic charge is equivalent to aligning the normal to the preferred foliation with the timelike Killing vector at the asymptotic boundary. Precisely in this aligned limit, the additional term in the variation of the mass vanishes.

The corresponding interpretation in Einstein--\ae ther theory has been less clear. This was puzzling because universal horizons also occur in Einstein--\ae ther theory, and in the static, spherically symmetric sector its black hole solution space is isomorphic to that of Ho\v{r}ava--Lifshitz gravity. Nevertheless, Einstein--\ae ther theory does not possess the khronon reparameterization symmetry, and therefore does not have the same elliptic charge.

In this work, we showed that, in the asymptotically AdS and spherically symmetric sector with $c_{14}=0$, Einstein--\ae ther theory instead possesses an additional \ae ther symmetry with its own conserved current and associated charge. We further showed that this charge becomes trivial at the asymptotic boundary precisely when the \ae ther aligns with the timelike Killing vector at infinity. Consequently, in the aligned limit, the \ae ther charge cannot contribute an additional term to the first law. This provides the Einstein--\ae ther interpretation of why, in the asymptotically aligned case, one obtains a coherent universal horizon first law containing only the expected entropy and enthalpy contributions.

Having established the relation between the additional term in the mass variation and the \ae ther charge, the next step is to treat this charge as an independent thermodynamic parameter. To obtain the full first law for asymptotically AdS universal horizons in Einstein--\ae ther theory, one must first express the mass as a function of the \ae ther charge and then vary the mass with respect to that charge. This is technically nontrivial, but conceptually necessary. The unexplained term in the mass variation should be interpreted as the contribution conjugate to the \ae ther charge.

The explicit charge-dependent form of the mass formula for asymptotically AdS universal horizons will be presented in a forthcoming paper, together with its variation and the resulting first law.

\acknowledgments

GN is supported by the 1st edition (2025) of ``Research fellowship in fundamental physics and study of the universe'' established by the Antonio Madonna Foundation ETS. GN is also grateful for the hospitality of Perimeter Institute where part of this work was carried out. We acknowledge Claude and ChatGPT for use in some algebraic calculations and as conversation partners.

\bibliographystyle{JHEP}
\bibliography{biblio}

\end{document}